\documentclass[11pt]{article}
\usepackage[square,sort&compress]{natbib}
\usepackage{natpatch}

\usepackage{epsfig}
\usepackage{graphicx}

\def\locald{\rho_0}

\def\mnucl{m_{N}}
\def\redN{\mu_N}
\def\redN{\mu_N}
\def\sigmaN{\sigma_{WN}}

\def\bra#1{\langle#1|}
\def\ket#1{|#1\rangle}
\def\vev#1{\langle#1\rangle}

\def\chione{\chi_1^{0}}
\def\bone{B_0^{(1)}}
\def\zone{Z^{(1)}}
\def\hone{H^{(1)}}

\newcommand{\mchi}{m_{\chi}}
\newcommand{\sigsip}{\sigma^{SI}_{p}}

\newcommand{\tev}{\,\mbox{TeV}}
\newcommand{\gev}{\,\mbox{GeV}}
\newcommand{\mev}{\,\mbox{MeV}}
\newcommand{\kev}{\,\mbox{keV}}

\newcommand{\pb}{\,\mbox{pb}}

\newcommand\lsim{\mathrel{\rlap{\lower4pt\hbox{\hskip1pt$\sim$}}
    \raise1pt\hbox{$<$}}}
\newcommand\gsim{\mathrel{\rlap{\lower4pt\hbox{\hskip1pt$\sim$}}
    \raise1pt\hbox{$>$}}}
\def\lsim{\mathrel{\raise.3ex\hbox{$<$\kern-.75em\lower1ex\hbox{$\sim
$}}}}
\def\gsim{\mathrel{\raise.3ex\hbox{$>$\kern-.75em\lower1ex\hbox{$\sim
$}}}}

\newcommand{\ie}{{\it i.e.}}

\newcommand{\s}{\,\mbox{s}}
\newcommand{\km}{\,\mbox{km}}
\newcommand{\kpc}{\,\mbox{kpc}}
\newcommand{\cm}{\,\mbox{cm}}

\begin{document}

\title{\huge{\bf Direct detection of WIMPs}
  \footnote{
    This contribution appeared as chapter 17, pp. 347-369, of 
    ``Particle Dark Matter: 
    Observations, Models and Searches'' edited by Gianfranco Bertone,
Copyright 2010 Cambridge University Press. Hardback ISBN 9780521763684,
http://cambridge.org/us/catalogue/catalogue.asp?isbn=9780521763684}}

\author{\Large{David G. Cerde\~no$^a$,
  Anne M. Green$^{b}$} \\[2ex] $^a$ Departamento de F\'{\i}sica Te\'{o}rica C-XI,
      and 
      Instituto de F\'{\i}sica Te\'{o}rica
      UAM-CSIC, \\[0pt] 
      Universidad Aut\'{o}noma de Madrid, 
      Cantoblanco, E-28049
      Madrid, Spain\\ 
      $^b$ School of Physics and Astronomy, 
      University of Nottingham\\[0pt]
      University Park, Nottingham, NG7 2RD, UK}  

\maketitle
\pagenumbering{arabic}

{\abstract{
    A generic weakly interacting massive particle (WIMP) is one of the
    most attractive candidates to account for the cold dark matter in
    our Universe, since it would be thermally
    produced with the correct abundance to account for the observed 
    dark matter density. 
    WIMPs can be searched for directly through their elastic
    scattering with a target material, and a variety of experiments 
    are currently operating or planned with this aim. 
    In these notes we overview the theoretical calculation of the
    direct detection rate of WIMPs as well as the different detection
    signals. We discuss the various ingredients (from particle physics
    and astrophysics) that enter the calculation and 
    review the theoretical
    predictions for the direct detection of WIMPs in particle
    physics models. 
} }

\section{Introduction}

If the Milky Way's DM halo is composed of WIMPs,
then the WIMP flux on the Earth is 
of order $10^{5} (100 \, {\rm GeV}/
m_{\chi}) \, {\rm cm}^{-2} \, {\rm s}^{-1}$.  This flux is sufficiently
large that, even though the WIMPs are weakly interacting, a small but
potentially measurable fraction will elastically scatter off
nuclei. Direct detection experiments aim to detect WIMPs via the
nuclear recoils, caused by WIMP elastic scattering, in dedicated
low background detectors~\cite{Goodman:1984dc}. More specifically they aim to
measure the rate, $R$, and energies, $E_{R}$, of the nuclear recoils
(and in directional experiments the directions as well).

In this chapter we overview the theoretical calculation of the direct
detection event rate and the potential direct detection signals.
Sec.~\ref{seceventrate} outlines the calculation of the event rate,
including the spin independent and dependent contributions and the
hadronic matrix elements. Sec.~\ref{astro} discusses the astrophysical
input into the event rate calculation, including the local WIMP
velocity distribution and density. In Sec.~\ref{signals} we describe
the direction detection signals, specifically the energy, time and
direction dependence of the event rate. Finally in
Sec.~\ref{secparticle} we discuss the predicted ranges for the WIMP mass
and cross-sections in various particle physics models.

\section{Event rate}
\label{seceventrate}

The differential event rate, usually expressed in terms of
counts/kg/day/keV (a quantity referred to as 
a differential rate unit or {\em dru})
for a WIMP with mass $\mchi$ and a nucleus with mass $\mnucl$
is given by
\begin{equation}
\label{drde}
  \frac{dR}{dE_R}=\frac{\locald}{\mnucl\,\mchi}\int_{v_{min}}^\infty  v
  f(v) \frac{d\sigmaN}{dE_R}(v,E_R)\, d v\,,
  \label{diff_rate}
\end{equation}
where $\rho_0$ is the local WIMP density,
$\frac{d\sigmaN}{dE_R}(v,E_R)$ is the differential cross-section for
the WIMP-nucleus elastic scattering and $f(v)$ is the WIMP speed
distribution in the detector frame normalized to unity.

Since the WIMP-nucleon relative speed is of order $100 \, {\rm
  km}^{-1} \, {\rm s}^{-1}$ the elastic scattering  occurs in
the extreme non-relativistic limit, and the recoil energy of the
nucleon is easily calculated in terms of the scattering angle in the
center of mass frame, $\theta^{*}$
\begin{equation}
  E_R=\frac{\redN^2v^2(1-\cos\theta^{*})}{\mnucl} \,,
\end{equation}
where $\redN=\mchi \mnucl/(\mchi + \mnucl)$ is the WIMP-nucleus reduced mass.

The lower limit of the integration over WIMP speeds is given by the
minimum WIMP speed which can cause a recoil of energy $E_{R}$:
$v_{min}=\sqrt{(\mnucl E_R)/(2\redN^2)}$. The upper limit is formally
infinite, however the local escape speed $v_{\rm esc}$ (see
Sec.~\ref{fv}), is the maximum speed {\em in the Galactic rest
  frame} for WIMPs which are gravitationally bound to the Milky Way.

The total event rate (per kilogram per day) is found by integrating
the differential event rate over all the possible recoil
energies: 
 \begin{equation}
  R=\int_{E_T}^\infty
  dE_R\frac{\locald}{\mnucl\,\mchi}\int_{v_{min}}^\infty  
  v f(v) \frac{d\sigmaN}{dE_R}(v,E_R)\, d v \,,
\end{equation}
where $E_{T}$ is the threshold energy, the smallest recoil energy
which the detector is capable of measuring.

The WIMP-nucleus differential cross section 
encodes the particle physics inputs (and associated
uncertainties) including the WIMP interaction properties.
It depends fundamentally on
the WIMP-quark interaction strength, which is calculated from the
microscopic description of the model, in terms of an effective
Lagrangian describing the interaction of the particular WIMP candidate
with quarks
and gluons. The resulting cross section is then promoted to a
WIMP-nucleon cross section. This entails the use of hadronic matrix
elements, which describe the nucleon content in quarks and gluons, and
are subject to large uncertainties. 
In general, 
the WIMP-nucleus cross section can be separated into a
spin-independent (scalar) and a spin-dependent contribution, 
\begin{equation}
  \frac{d\sigmaN}{dE_R}= \left(\frac{d\sigmaN}{dE_R}\right)_{SI}
  +\left(\frac{d\sigmaN}{dE_R}\right)_{SD}\,.
\end{equation}
Finally, the total WIMP-nucleus cross section is calculated by adding
coherently the above spin and scalar components, using nuclear wave
functions. The form factor, $F(E_R)$, encodes the
dependence on the momentum transfer, $q=\sqrt{2 \mnucl E_R}$, 
and accounts for the
coherence loss which leads to a suppression in the event rate
for heavy WIMPs or nucleons. 
In general, we can express the differential cross section as
\begin{equation}
  \frac{d\sigmaN}{dE_R}=\frac{\mnucl}{2\redN^2v^2}\left(\sigma_0^{SI}
  F^2_{SI}(E_R) + \sigma_0^{SD} F^2_{SD}(E_R)\right)\ ,
\end{equation}
where $\sigma_0^{SI,\, SD}$ are the spin-independent and -dependent cross
sections at zero momentum transfer. 

The origin of the different contributions is best understood at the
microscopic level, by analysing the Lagrangian which describes the
WIMP interactions with quarks. The contributions to the
spin-independent cross section arise from scalar and vector
couplings to quarks, whereas the spin-dependent part of the
cross section originates from axial-vector couplings.
These contributions are characteristic of the particular WIMP candidate
(see, e.g., \cite{Barger:2008qd}) and can be potentially useful for
their discrimination in direct detection experiments.

\subsection{Spin-dependent contribution}

The contributions to the spin-dependent (SD) part of the WIMP-nucleus
scattering cross section arise from couplings of the WIMP field 
to the quark axial current,
$\bar q\gamma_\mu\gamma_5 q$. For example, if the WIMP is a (Dirac or 
Majorana) fermion, such as the lightest neutralino in supersymmetric
models, the Lagrangian can contain the term
\begin{equation}
  {\cal L}\supset \alpha_q^A (\bar\chi\gamma^\mu\gamma_5\chi)
  (\bar q\gamma_\mu\gamma_5 q) \,.
\end{equation}
If the WIMP is a spin 1 field, such as in the case of LKP and LTP, 
the interaction term is slightly
different, 
\begin{equation}
  {\cal L}\supset \alpha_q^A
  \epsilon^{\mu\nu\rho\sigma}
  (B_\rho\stackrel{\leftrightarrow}{\partial_\mu} B_\nu) 
  (\bar q\gamma^\sigma\gamma_5 q) \,.
\end{equation}

In both cases, the nucleus, $N$, matrix element reads 
\begin{equation}
  \bra{N}\bar q\gamma_\mu\gamma_5q
  \ket{N}
  =2\lambda_q^N\bra{N}J_N\ket{N} \,,
\end{equation}
where the coefficients 
$\lambda_q^N$ relate the quark spin matrix elements to the
angular momentum of the nucleons. They can be parametrized as 
\begin{equation}
  \lambda_q^N\simeq\frac{\Delta_q^{(p)}\vev{S_p}
    +\Delta_q^{(n)}\vev{S_n}}{J} \,,
\end{equation}
where $J$ is the total angular momentum of the nucleus, 
the quantities $\Delta q^{n}$ are related to the matrix element
of the axial-vector current in a nucleon,
  $\bra{n}\bar q\gamma_\mu\gamma_5q\ket{n}=2s_\mu^{(n)}
  \Delta_q^{(n)}$,
and $\vev{S_{p,n}}=\bra{N}S_{p,n}\ket{N}$
is the expectation value of the 
spin content of the proton or neutron group in the
nucleus\footnote{These quantities can be determined from simple
  nuclear models. For example, the single-particle shell model assumes
  the nuclear spin is solely due to the spin of the single unpaired
  proton or neutron, and therefore vanishes for even nuclei. More
  accurate results can be obtained by using detailed nuclear
  calculations.}.
Adding the contributions from the different quarks, it is customary to
define 
\begin{equation}
  a_p=\sum_{q=u,d,s}\frac{\alpha_q^A}{\sqrt{2}G_F}\Delta_q^{p}\,;\quad
  a_n=\sum_{q=u,d,s}\frac{\alpha_q^A}{\sqrt{2}G_F}\Delta_q^{n} \,,
\end{equation}
and 
\begin{equation}
  \Lambda=\frac1J\left[a_p\langle S_p\rangle+a_n\langle
    S_n\rangle\right]  \,.
\end{equation}
The resulting differential cross section can then be expressed (in the
case of a fermionic WIMP) as
\begin{equation}
  \left(\frac{d\sigmaN}{dE_R}\right)_{SD}=\frac{16 \mnucl}{\pi
    v^2}\Lambda^2 G_F^2 J(J+1) \frac{S(E_R)}{S(0)} \,,
\end{equation}
(using $d|\vec q|^2=2\mnucl dE_R$). 
The expression for a spin 1 WIMP can be found, e.g., in
Ref.~\cite{Barger:2008qd}.

In the parametrization of the form factor
it is common to use a decomposition into isoscalar, $a_0=a_p+a_n$, and  
isovector, $a_1=a_p-a_n$, couplings 
\begin{equation}
  S(q)=a_0^2 S_{00}(q)+a_0a_1S_{01}(q)+a_1^2S_{11}(q) \,,
\end{equation}
where the parameters $S_{ij}$ are determined experimentally.

\subsection{Spin-independent contribution}

Spin-independent (SI) contributions to the total cross section may arise
from scalar-scalar and vector-vector couplings in the Lagrangian:
\begin{equation}
  {\cal L}\supset \alpha_q^S
  \bar\chi\chi \bar qq +
  \alpha_q^V\bar\chi\gamma_\mu\chi\bar q\gamma^\mu q
  \,.
\end{equation}
The
presence of these couplings depends on the particle physics model
underlying the WIMP candidate.
In general one can write 
\begin{equation}
  \left(\frac{d\sigmaN}{dE_R}\right)_{SI}=\frac{\mnucl \sigma_0
  F^2(E_R) }{2\redN^2v^2} \,,
\end{equation}
where the nuclear form factor for coherent interactions $F^2(E_R)$ 
can be qualitatively understood as a Fourier transform of the nucleon
density and is
usually parametrized in terms of the momentum transfer 
as \cite{Helm:1956zz,Engel:1991wq}
\begin{equation}
  F^2(q)=\left(\frac{3j_1(qR_1)}{qR_1}\right)^2\,
  \exp\left[-q^2s^2\right] \,,
\end{equation}
where $j_{1}$ is a spherical Bessel function, $s\simeq 1$~fm is a
measure of the nuclear skin thickness, and $R_1=\sqrt{R^2-5s^2}$ with
$R\simeq 1.2\, A^{1/2} \, {\rm fm}$.  The form factor is normalized to
unity at zero momentum transfer, $F(0)=1$.

The contribution from the scalar coupling leads to the following
expression for the WIMP-nucleon cross section, 
\begin{equation}
  \sigma_0=\frac{4 \redN^2}{\pi} \left[Z f^p + (A-Z) f^n\right]^2 \,,
\end{equation}
with
\begin{equation}
  \frac{f^p}{m_p}=\sum_{q=u,d,s}\frac{\alpha^S_q}{m_q} f_{Tq}^p + 
  \frac{2}{27} f_{TG}^p\sum_{q=c,b,t}\frac{\alpha^S_q}{m_q} \,,
\end{equation}
where the quantities $f_{Tq}^{p}$ represent the contributions of the light
quarks to the mass of the proton, and are defined as
$m_pf_{Tq}^p\equiv\langle p|m_q\bar qq|p\rangle$. Similarly the second
term is due to the interaction of the WIMP and the gluon
scalar density in the nucleon, with 
$f_{TG}^{p}=1-\sum_{q=u,d,s}f_{Tq}^{p}$. They are
determined experimentally,
\begin{equation}
  f_{Tu}^p=0.020\pm 0.004,\quad
  f_{Td}^p=0.026\pm 0.005,\quad
  f_{Ts}^p=0.118\pm 0.062,
  \label{si_scalar_densities}
\end{equation}
with  $f_{Tu}^n=f_{Td}^p$, $f_{Td}^n=f_{Tu}^p$, and 
$f_{Ts}^n=f_{Ts}^p$.
The uncertainties in these quantities, among which the most important
is that on $f_{Ts}$, mainly 
stem from the determination of the $\pi$-nucleon sigma term.

The vector coupling (which is present, for example, in the case of a
Dirac fermion but vanishes for Majorana particles)
gives rise to an extra contribution.
Interestingly, the sea quarks and gluons do not contribute to
the vector current. Only valence quarks contribute, leading to the
following expression
\begin{equation}
  \sigma_0=\frac{\redN^2 B_N^2}{64\pi} \,,
\end{equation}
with 
\begin{equation}
  B_N\equiv \alpha_u^V(A+Z) + \alpha_d^V (2A-Z)
     \,.
\end{equation}

Thus, for a general WIMP with both scalar and vector interactions, the
spin-independent contribution to the scattering cross section would
read 
\begin{equation}
   \left(\frac{d\sigmaN}{dE_R}\right)_{SI}=\frac{2\,\mnucl}{\pi
    v^2}\left[\left[Z f^p + (A-Z) f^n\right]^2 
     + \frac{B_N^2}{256}\right]
   F^2(E_R) \,.
\end{equation}

In most cases the WIMP coupling to neutrons and protons is very
similar, $f^p\approx f^n$, and therefore the scalar contribution can be
approximated by 
\begin{equation}
   \left(\frac{d\sigmaN}{dE_R}\right)_{SI}=\frac{2\,\mnucl\,A^2
     (f^p)^2}{\pi v^2}  
   F^2(E_R) \,.
\end{equation}

The spin-independent contribution basically scales as the square of
the number of nucleons ($A^2$), whereas the spin-dependent one is
proportional to a function of the nuclear angular momentum, $(J+1)/J$.
Although in general both have to be taken into account, 
the scalar component dominates for heavy targets ($A> 20$), which is the
case for most experiments (usually based on targets with heavy nuclei 
such as Silicon, Germanium, Iodine or Xenon).
Nevertheless, dedicated experiments exist that are also sensitive to
the SD WIMP coupling through the choice of targets with a
large nuclear angular momentum.

As we have seen, the WIMP direct detection rate depends on both
astrophysical input (the local DM density and velocity distribution,
in the lab frame) and particle physics input (nuclear form factors and
interaction cross-sections, which depend on the theoretical framework
in which the WIMP candidate arises).  We will discuss these inputs in
more detail in Secs.~\ref{astro} and \ref{secparticle} respectively.

\subsection{Hadronic Matrix Elements}

The effect of uncertainties in the hadronic matrix elements has
been studied in detail for the specific case of neutralino dark matter 
\cite{Bottino:1999ei,Ellis:2000ds,Bottino:2001dj,Ellis:2005mb,Ellis:2008hf}.  
Concerning the SI cross section, 
the quantities $f_{Tq}^{p}$ in Eq.(\ref{si_scalar_densities}) can be
parametrized in terms of the $\pi$ nucleon sigma term, $\Sigma_{\pi
  N}$, (see in this respect, e.g., Refs.\cite{Bottino:2001dj,Ellis:2008hf}) 
which, in terms of the $u$
and $d$ quark masses reads
\begin{equation}
  \Sigma_{\pi N}=\frac12\left(m_u+m_d\right)
  \langle N|\bar uu+\bar dd|N\rangle\ ,
\end{equation}
and is related to the
strange quark scalar density in the nucleon. 
The largest source of uncertainty in $f_{Tq}^{p}$ 
stems from the determination of this quantity, for which the current data
implies $\Sigma_{\pi N}=(64\pm8)\mev$ \cite{Pavan:2001wz},
which translates into a variation of a factor $4$ in $f_{Ts}$.
Notice that in general the WIMP interaction with strange quarks would
be the leading contribution to the SI cross section, due to its larger
Yukawa coupling. 
In this case, $\sigma_0$ is roughly proportional to $f_{Ts}\,^2$ and
the above uncertainty in the strange quark content leads to a
variation of more than one order of magnitude in the resulting SI 
cross section 
\cite{Bottino:2001dj,Ellis:2008hf}).

Similarly, for the SD cross section the uncertainties in the strange
spin contribution $\Delta_s$ are the dominant contribution to the
error in $\sigma_0$. 
However, in the case of the neutralino, this 
can imply a correction of as much as a factor 2 
in the resulting cross section
\cite{Ellis:2008hf}, being therefore much smaller than the above
uncertainty for the SI cross section.
It should be emphasized, however, that 
uncertainties in the determination of the spin form factors $S(q)$
would also affect the theoretical predictions for the dark matter
detection rate.

\section{Astrophysics input }
\label{astro}

\subsection{Local DM density}
\label{dmdensity}

The differential event rate is directly proportional to the local WIMP
density, $\rho_{0} \equiv \rho(r=R_{0})$ where $R_{0}= (8.0 \pm 0.5)
\kpc $~\cite{Reid:1993fx} is the solar radius. Any observational
uncertainty in $\rho_{0}$ therefore translates directly into in an
uncertainty in the event rate and the inferred constraints on, or
measurements of, the scattering cross-sections.

Exclusion limits are traditionally calculated assuming a canonical
local WIMP density, $\rho_{0} = 0.3 \gev \cm^{-3}$.  The local WIMP
density is calculated by applying observational constraints (including
measurements of the rotation curve) to models of the Milky Way and the
values obtained can vary by a factor or order 2 depending on the models used~\cite{Caldwell:1981rj,1983ApJ...265..730B,Gates:1995dw,Bergstrom:1997fj}.
A recent study~\cite{Widrow:2008yg}, using spherical
halo models with a cusp ($\rho \propto r(r)^{-\alpha}$ as $r \rightarrow 0$)
finds $\rho_{0} = (0.30 \pm 0.05) \gev \cm^{-3}$.

\subsection{Speed distribution}
\label{fv}

The standard halo model, conventionally used in calculations of
exclusion limits and signals, has an isotropic, Gaussian velocity
distribution (often referred to as Maxwellian)
\begin{equation}
f({\bf v}) = \frac{1}{ \sqrt{2 \pi} \sigma} \exp{\left( -\frac{|{\bf v}|^2}
 {2 \sigma^2} \right)}\,.
\end{equation}
The speed dispersion is related to the local circular speed by $\sigma
= \sqrt{3/2} v_{c}$ and  $v_{\rm c}= (220 \pm 20) \km
\s^{-1}$~\cite{Kerr:1986hz} (see Sec.\ref{earthmotion}) so that
$\sigma \approx 270 \km \s^{-1}$.  This velocity distribution
corresponds to an isotropic singular isothermal sphere with density
profile $\rho(r) \propto r^{-2}$. The isothermal sphere is simple, and
not unreasonable as a first approximation, however it is unlikely to
be an accurate model of the actual density and velocity distribution of
the Milky Way. Observations and numerical simulations (see chapter 1)
indicate that dark matter halos do not have a $1/r^2$ density profile
and are (to some extent at least) triaxial and anisotropic.

If the velocity distribution is isotropic
there is a one to one relation between $f({\bf v})$ and the the
spherically symmetric density profile given by Eddington's
formula~\cite{1916MNRAS..76..572E}, see
Refs.~\cite{Ullio:2000bf,Vergados:2002hc}. In general the steady state
phase-space distribution of a collection of collisionless particles is
given by the collisionless Boltzmann equation and the velocity
dispersions of the system are calculated via the Jean's equations
(e.g. Ref.~\cite{1987gady.book.....B}). Solving the Jean's equations
requires assumptions to be made, and therefore even for a specific
density distribution the solution is not unique. Several specific
models have been used in the context of WIMP direct detection
signals.  The logarithmic ellipsoidal model~\cite{Evans:2000gr} is the
simplest triaxial generalisation of the isothermal sphere and has a
velocity distribution which is a multi-variate
Gaussian. Osipkov-Merritt models~\cite{Osipkov,1985AJ.....90.1027M}
are spherically symmetric with radially dependent anisotropic
velocity distributions. Fitting functions for the speed distributions in
these models are available, for a selection of density profiles, in
Ref.~\cite{2000ApJS..131...39W}. 
Velocity distributions have also been extracted from cosmological
simulations, with both multi-variate Gaussian~\cite{Helmi:2002ss} and
Tsallis~\cite{Tsallis:1987eu} distributions~\cite{Hansen:2005yj} being
advocated as fitting functions. 
While it is not known whether any
of these models provide a good approximation to the real local
velocity distribution function, the models are none the less useful for
assessing the uncertainties in the direct detection signals.

Particles with speed, in the Galactic rest frame, greater than the
local escape speed, $v_{esc}= \sqrt{2 |\Phi(R_{0})|}$ where $\Phi(r)$ is the
potential, are not gravitationally bound. Many of the models used, in
particular the standard halo model, formally extend to infinite radii
and therefore their speed distribution has to be truncated at
$v_{esc}$ `by hand' (see e.g. Ref.~\cite{Drukier:1986tm}).  The
standard value for the escape speed is $v_{esc} = 650 \km \s^{-1}$.  A recent
analysis, using high velocity stars from the RAVE survey, finds $498
{\rm km \, s}^{-1} < v_{esc} < 608 \km \s^{-1}$ with a median
likelihood of $544 \km \s^{-1}$~\cite{Smith:2006ym}.

In Sec.~\ref{signals} we discuss the impact of uncertainty in the speed
distribution on the direct detection signals.

\subsection{Earth's motion}
\label{earthmotion}

The WIMP speed distribution in the detector rest frame is calculated
by carrying out, a time dependent, Galilean transformation: ${\bf v}
\rightarrow \tilde{\bf{v}} = {\bf v} + {\bf v}_{e}(t)$.  The Earth's
motion relative to the Galactic rest frame, ${\bf v}_{e}(t)$, is made
up of three components: the motion of the Local Standard of Rest
(LSR), the Sun's peculiar motion with respect to the LSR, ${\bf
  v}_{\odot}^{\rm p}$, and the Earth's orbit about the Sun, ${\bf
  v}_{e}^{\rm orb}$ .

If the Milky Way is axisymmetric then the motion of the LSR is given
by the local circular velocity $(0,v_{c},0)$, where $v_{\rm c}= 220
\km \s^{-1}$ is the standard value. Kerr and Lynden-Bell found, by
combining a large number of independent measurements, $v_{\rm c}= (222
\pm 20) \km \s^{-1}$~\cite{Kerr:1986hz}. A more recent determination,
using the proper motions of Cepheids measured by
Hipparcos~\cite{1997MNRAS.291..683F}, is broadly consistent: $v_{c}=
(218 \pm 7) \km \s^{-1}(R_{0}/8 \kpc)$.

The Sun's peculiar motion, determined using the parallaxes and proper
motions of stars in the solar neighbourhood from the Hipparcos
catalogue, is $ {\bf v}_{\odot}^{\rm p} = (10.0 \pm 0.4, 5.2 \pm 0.6, 7.2
\pm 0.4) \km \s^{-1}$~\cite{Dehnen:1997cq} in Galactic
co-ordinates (where $x$ points towards the Galactic center, $y$ is the
direction of Galactic rotation and $z$ towards the North Galactic
Pole).

A relatively simple, and reasonably accurate, expression for the
Earth's motion about the Sun can be found by ignoring the ellipticity
of the Earth's orbit and the non-uniform motion of the Sun in right
ascension~\cite{Gelmini:2000dm}: ${\bf v}_{e}^{\rm orb} = v_{e} [{\bf
  e}_{1} \sin{\lambda(t)} - {\bf e}_{2} \cos{\lambda(t)}]$ where $v_{e}=
29.8 \km \s^{-1}$ is the orbital speed of the Earth,
$\lambda(t)=2 \pi (t-0.218)$ is the Sun's ecliptic longitude (with $t$ in
years) and ${\bf e}_{1(2)}$ are unit vectors in the direction of the
Sun at the Spring equinox (Summer solstice). In Galactic co-ordinates
${\bf e}_{1}= (-0.0670, 0.4927, -0.8676)$ and ${\bf e}_{2}=(-0.9931,
-0.1170, 0.01032)$. More accurate expressions can be found in
Ref.~\cite{1985spas.book.....G}.

The main characteristics of the WIMP signals can be found using only
the motion of the LSR, and for the time dependence the component of the Earth's
orbital velocity in that direction.  However accurate calculations, for
instance for comparison with data, require all the components
described above to be taken into account.

\subsection{Ultra-fine structure}
\label{ultra}
Most of the WIMP velocity distributions discussed in Sec.~\ref{fv} are
derived by solving the collisionless Boltzmann equation, which assumes
that the phase space distribution has reached a steady state. However
this may not be a good assumption for the Milky Way; structure
formation in CDM cosmologies occurs hierarchically and the relevant
dynamical timescales for the Milky Way are not many orders of
magnitude smaller than the age of the Universe.

Both astronomical observations and numerical simulations (due to their
finite resolution) typically probe the dark matter distribution on
$\sim \kpc$ scales. Direct detection experiments probe the DM
distribution on sub milli-pc scales (the Earth's speed with respect to
the Galactic rest frame is $\approx 0.2 \, {\rm mpc} \, {\rm
  yr}^{-1})$. It has been argued (see
e.g. Refs.~\cite{Moore:2001vq,Stiff:2003tx,Fantin:2008ur}) that on
these scales the DM may not have yet reached a steady state and could
have a non-smooth phase-space distribution.
On the other hand it has been argued that the rapid decrease in
density of streams evolving in a realistic, ellipsoidal, Galactic
potential means that there will be a large number of overlapping
streams in the solar neighbourhood producing an effectively smooth
DM distribution~\cite{Vogelsberger:2007ny}. 

If the local DM distribution consists of a small number of streams,
rather than a smooth distribution, then there would be significant
changes in the signals which we will discuss in Sec.~\ref{signals}.  This is 
currently an open issue; directly calculating the DM distribution on
the scales probed by direct detection experiments is a difficult and
unresolved problem.

It has been suggested that a tidal stream from the Sagittarius (Sgr)
dwarf galaxy, which is in the process of being disrupted, passes
through the solar neighbourhood with the associated DM potentially
producing
distinctive signals in direct detection
experiments~\cite{Freese:2003tt,Freese:2003na}. Subsequent numerical
simulations of the disruption of Sgr along with observational searches
for local streams of stars suggest that the Sgr stream does not in
fact pass through the solar neighbourhood (see
e.g. Ref.~\cite{Seabroke:2007er}).  None the less the calculations of
the resulting WIMP signals in Refs.~\cite{Freese:2003tt,Freese:2003na}
provide a useful illustration of the qualitative effects of streams.

\section{Signals}
\label{signals}

We have already seen in Sec.~\ref{seceventrate} that the recoil rate is energy
dependent due to the kinematics of elastic scattering, combined with
the WIMP speed distribution. Due to the motion of the Earth with
respect to the Galactic rest frame the recoil rate is also both time
and direction dependent.  In this section we examine the energy, time
and direction dependence of the recoil rate and the resulting WIMP
signals. In each case we first focus on the signal expected for the
standard halo model, with a Maxwellian velocity distribution, before
discussing the effect on the signal of changes in the WIMP velocity
distribution.

\subsection{Energy dependence}
\label{sec:drde}

The shape of the differential event rate depends on the WIMP and
target masses, the WIMP velocity distribution and the form factor. For 
the standard halo model the expression for the 
differential event rate, eq.~\ref{drde}, can be rewritten approximately
 (c.f. Ref.\cite{Lewin:1995rx}) as
\begin{equation}
\label{drdeapprox}
\frac{{\rm d} R}{{\rm d} E_{R}} \approx \left( \frac{{\rm d} R}{{\rm d}E_{R}}
   \right)_{0} F^2(E_{R}) \exp{ \left(- \frac{E_{R}}{E_{c}} \right)}  \,,
\end{equation}
where $({\rm d} R/ {\rm d} E_{R})_{0}$ is the event rate in the $E
\rightarrow 0$ keV limit.  The characteristic energy scale
is given by $E_{c} = (c_{1}2 \mu_{N}^2 v_{c}^2)/m_{N}$ where $c_{1}$
is a parameter of order unity which depends on the target nuclei. If the
WIMP is much lighter than the target nuclei, $\mchi \ll m_{N}$, then
$E_{c} \propto \mchi^2 / m_{N} $ while if the WIMP is much heavier
than the target nuclei $E_{c} \propto m_{N}$. The total recoil rate is
directly proportional to the WIMP number density, which varies as $1/\mchi$. 

\begin{figure}[!t]
\centering
\includegraphics[width=4.0in,angle=0]{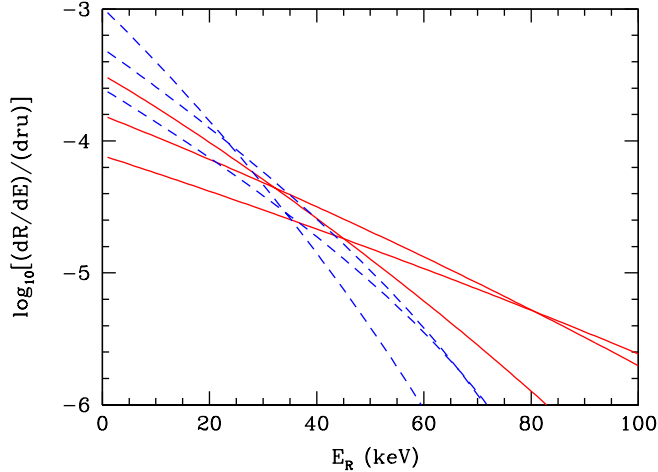}
\vspace*{-1.5cm}
\caption{The dependence of the spin independent differential event rate on the 
 WIMP mass and target. The solid and dashed
  lines are for Ge and Xe respectively and WIMP masses of (from top
to bottom at $E_{R}=0 \kev$) 50, 100 and 200 keV. The scattering 
cross-section on the proton is taken to be
$\sigsip=10^{-8} \pb$.}
\label{fig_drde}
\end{figure}

In fig.~\ref{fig_drde} we plot the differential event rate for Ge and
Xe targets and a range of WIMP masses. As expected, for a fixed target
the differential event rate decreases more rapidly with increasing
recoil energy for light WIMPs. For a fixed WIMP mass the decline of
the differential event rate is steepest for heavy target nuclei. The
dependence of the energy spectrum on the WIMP mass allows the WIMP
mass to be estimated from the energies of detected events
(e.g. Ref.~\cite{Green:2007rb}). Furthermore the consistency of energy
spectra measured by experiments using different target nuclei would
confirm that the events were due to WIMP scattering (rather than, for
instance, neutron backgrounds)~\cite{Lewin:1995rx}. In particular, for
spin independent interactions, the total event rate scales as $A^2$.
The is sometimes referred to as the `materials signal'.

The WIMP and target mass dependence of the differential event rate
also have some general consequences for experiments. The
dependence of the total event rate on $\mchi$ means that, for fixed
cross-section, a larger target mass will be required to detect heavy
WIMPs than lighter WIMPs. For very light WIMPs the rapid
decrease of the energy spectrum with increasing recoil energy means
that the event rate above the detector threshold energy, $E_{T}$, may
be small. If the WIMP is light, $ < {\cal O}(10 \gev)$, a detector
with a low, $<{\cal O}(\kev)$, threshold energy will be required.

The most significant astrophysical uncertainties in the differential
event rate come from the uncertainties in the local WIMP density and
circular velocity. As discussed in Sec.~\ref{dmdensity} the
uncertainty in the local DM density translates directly into an
uncertainty in constraints on (or in the future measurements of) the
scattering cross-section. The {\em time averaged} differential event
rate is found by integrating the WIMP velocity distribution, therefore
it is only weakly sensitive to changes in the shape of the WIMP
velocity distribution. For the smooth halo models discussed in
Sec.~\ref{fv} the {\em time averaged} differential event rates are
fairly similar to that produced by the standard halo
model~\cite{Kamionkowski:1997xg,Donato:1998pc}. Consequently exclusion
limits vary only weakly~\cite{Donato:1998pc,Green:2002ht} and there
would be a small (of order a few per-cent) systematic uncertainty in
the WIMP mass deduced from a measured energy
spectrum~\cite{Green:2008rd}. With multiple detectors it would in principle
be possible to measure the WIMP mass without any assumptions about
the WIMP velocity distribution~\cite{Drees:2008bv}.

In the extreme case of the WIMP distribution being composed of a small
number of streams the differential event rate would consists of a
series of (sloping due to the form factor) steps. The positions of the
steps would depend on the stream velocities and the target and WIMP
masses, while the relative heights of the steps would depend on the stream
densities.

\subsection{Time dependence}

The Earth's orbit about the Sun leads to a time dependence,
specifically an annual modulation, in the differential event
rate~\cite{Drukier:1986tm,Freese:1987wu}. The Earth's speed with
respect to the Galactic rest frame is largest in Summer when the
component of the Earth's orbital velocity in the direction of solar
motion is largest. Therefore the number of WIMPs with high (low)
speeds in the detector rest frame is largest (smallest) in
Summer. Consequently the differential event rate has an annual
modulation, with a peak in Winter for small recoil energies and in
Summer for larger recoil energies~\cite{Primack:1988zm}. The energy at
which the annual modulation changes phase is often referred to as the
`crossing energy'.

\begin{figure}
\centering
\includegraphics[width=4.0in,angle=0]{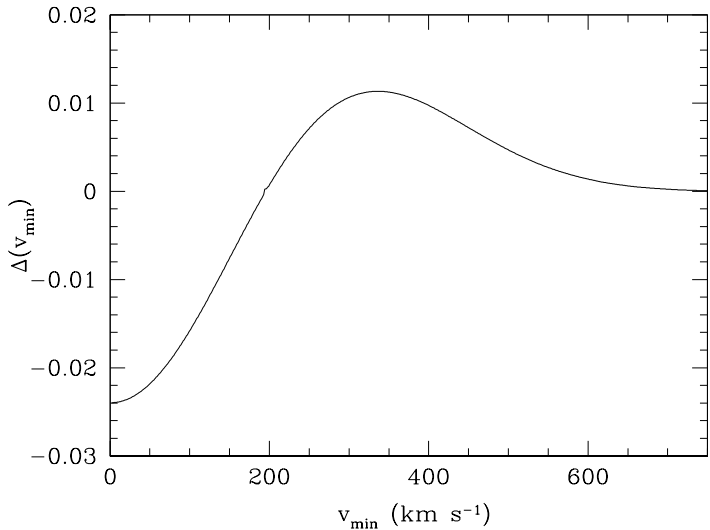}
\caption{Dependence of the amplitude of the annual modulation,
  $\Delta(v_{\rm min})$, on $v_{\rm min}$.}
\label{fig_am}
\end{figure}

Since the Earth's orbital speed is significantly
smaller than the Sun's circular speed the amplitude of the modulation
is small and, to a first approximation, the differential event rate
can, for the standard halo model, be written approximately as a Taylor
series:
\begin{equation}
\frac{{\rm d} R}{{\rm d} E_{R}} \approx 
   \bar{\left(\frac{{\rm d} R}{{\rm d} E_{R}} \right)}
  \left[ 1 + \Delta(E_{R}) 
  \cos{\alpha(t)} \right] \,,
\end{equation}
where $\alpha(t) = 2 \pi (t-t_{0})/T$, $T=1$ year and $t_{0} \sim 150$
days.  In fig.~\ref{fig_am} we plot the energy dependence of the
amplitude in terms of $v_{\rm min}$ (recall that $v_{min} \propto
E_{R}^{1/2}$ with the constant of proportionality depending on the
WIMP and target nuclei masses). The amplitude of the modulation is of
order 1-10 \%.

The Earth's rotation provides another potential time dependence in the
form of a diurnal modulation as the Earth acts as a shield in front of
the detector~\cite{Collar:1992qc,Hasenbalg:1997hs}, however the
amplitude of this effect is expected to be small, $<
1\%$~\cite{Hasenbalg:1997hs}.

There has been a substantial amount of work on the annual modulation
for the non-Maxwellian velocity distributions described in Sec.~\ref{fv}
~\cite{Brhlik:1999tt,Belli:1999nz,Vergados:1999sf,Ullio:2000bf,Green:2000jg,Vergados:2002hc,Green:2003yh,Vergados:2007nc,Fairbairn:2008gz}.
In contrast to the
time-averaged differential event rate, both the phase and amplitude of
the annual modulation can vary substantially.  Consequently the
regions of WIMP mass-cross-section parameter space consistent with an
observed signal can change
significantly~\cite{Belli:1999nz,Belli:2002yt}. Note that if the
components of the Earth's orbital velocity perpendicular to the direction of
Solar motion are neglected, then the phase change will be
missed~\cite{Green:2003yh}. For a WIMP stream the position and height
of the step in the energy spectrum would vary annually
(e.g. Ref.~\cite{Savage:2006qr}).

\subsection{Direction dependence}	 

The detector motion with respect to the Galactic rest frame also
produces a directional signal. The WIMP flux in the lab frame is
sharply peaked in the direction of motion of the Earth, and hence the
recoil spectrum is also peaked in this direction (albeit less sharply
due to the elastic scattering).

The directional recoil rate is most
compactly written as~\cite{Gondolo:2002np}
\begin{equation}
\frac{{\rm d}R}{{\rm d} E_{R} {\rm d} \Omega} = 
\frac{\rho_{0} \sigma_{0} A^2}{4 \pi \mu_{p}^2 \mchi} F^2(E_{R}) 
   \hat{f}(v_{\rm min},\hat{{\bf q}}) \,,
\end{equation}
where ${\rm d} \Omega = {\rm d} \phi \, {\rm d} \cos{\theta}$,
$\hat{\bf{q}}$ is the recoil direction and
$\hat{f}(v_{\rm min},\hat{{\bf q}})$ is the 3-dimensional Radon transform
of the WIMP velocity distribution $f({\bf v})$
\begin{equation}
\hat{f}(v_{\rm min},\hat{{\bf q}}) = \int \delta({\bf v}.\hat{{\bf q}} - v_{\rm min}) f({\bf v})
 {\rm d}^3 v \,.
\end{equation}
Geometrically
the Radon transform is the integral of the function $f({\bf v})$ on a plane
orthogonal to the direction $\hat{\bf{q}}$ at a distance $v_{\rm min}$ from the origin.
See Ref.~\cite{Copi:2000tv} for an alternative, but equivalent, expression.

\begin{figure}
\centering
\includegraphics[width=4.0in,angle=0]{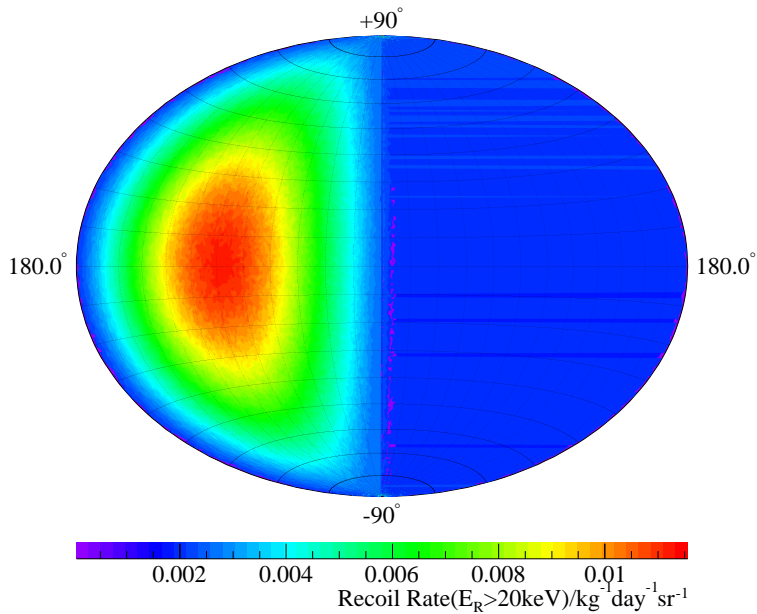}
\caption{Hammer-Aitoff projection of the directional
recoil rate for $S$, above $E_{\rm T}= \, 20 \, {\rm keV}$, due to a standard
halo of WIMPs with $\mchi= \, 100 \, {\rm GeV}$. This figure was generated
using the HADES directional dark matter simulation code written by Ben Morgan.}
\label{fig_dd}
\end{figure}

For the standard halo model the direction dependence
is approximately given by~\cite{Spergel:1987kx}
\begin{equation}
\frac{{\rm d} R}{{\rm d} E_{R} \, {\rm d} \cos{\gamma}}
   \propto \exp{\left[- \frac{(v_{\odot} \cos{\gamma} - v_{min})^2}{v_{c}^2}
  \right]} \,,
\end{equation}
where $\gamma$ is the angle between the recoil and the mean direction
of solar motion.  The distribution of recoil directions peaks in the
mean direction of motion of the Sun (towards the constellation
CYGNUS~\cite{SnowdenIfft:1999hz}) with the event rate in the forward
direction being roughly an order of magnitude larger than that in the
backward direction~\cite{Spergel:1987kx}, since the Sun's speed is
comparable to the mean WIMP speed. The directional recoil rate of
$\mchi=100$ GeV WIMPs on $S$ is shown in fig.~\ref{fig_dd}.

With an ideal directional detector, with 3-d read-out and capable of
measuring the senses of the recoils (\ie \, distinguishing between the
head and tail of each recoil), it would be possible to distinguish a
WIMP signal from isotropic backgrounds with only of order 10
events~\cite{Copi:1999pw,Copi:2000tv,Morgan:2004ys}. This number
increases significantly (by roughly an order of magnitude) if either
the senses can not be measured or the read out is only
2-d~\cite{Morgan:2005sq,Green:2006cb,Copi:2005ya}. Another potential
directional signal is the rotation of the mean recoil direction in the
lab over a sidereal day due to the motion of the
Earth~\cite{SnowdenIfft:1999hz}.  See chapter~22
for
the principles and practice of directional detection experiments.

For plausible smooth halo models changes in the WIMP velocity
distribution affect the detailed angular recoil rate. However the
rear-front asymmetry is robust and the number of events required by an
ideal detector only varies by of order
10\%~\cite{Copi:1999pw,Copi:2000tv,Morgan:2004ys}. For non-ideal
detectors the variation in the number of events required can be
larger~\cite{Morgan:2004ys,Morgan:2005sq}. With a large
number of events (of order thousands) it would be possible to probe
the detailed WIMP velocity
distribution~\cite{Morgan:2004ys,Morgan:2005sq,Host:2007fq}. A stream of WIMPs
produces a recoil spectrum which is peaked in the opposite direction
(e.g. Ref.~\cite{Alenazi:2007sy}).

\section{Particle Physics input} 
\label{secparticle}

Let us finally address the Particle Physics input to the determination
of the WIMP detection rate, which enters through the theoretical
predictions to the WIMP-nucleus scattering cross section. These are
sensitive to the WIMP nature.
We will here briefly summarise the results for WIMP candidates
that arise in various well-motivated theories of physics beyond the
Standard Model at the TeV scale (Supersymmetric theories, 
Universal Extra Dimension scenarios and Little Higgs models), as well as
phenomenologically motivated scenarios.

\subsection{Supersymmetric WIMPs}

The canonical and best studied supersymmetric WIMP is the lightest
neutralino, $\chione$.
Its detection properties are very dependent on its
composition.
More specifically, within the MSSM framework, 
in the expressions for the scattering amplitudes
\cite{Srednicki:1989kj,Gelmini:1990je,Drees:1992rr,Drees:1993bu,Ellis:2000ds},
the SI
part of the neutralino-nucleon cross section
receives contributions from Higgs
exchange in a $t$-channel and squark exchange in an $s$-channel. The
latter also contributes to the SD part of the cross
section,
together with a $Z$ boson exchange in a $t$-channel.

\begin{figure}[!t]
\centering
\includegraphics[width=3.1in,angle=0]{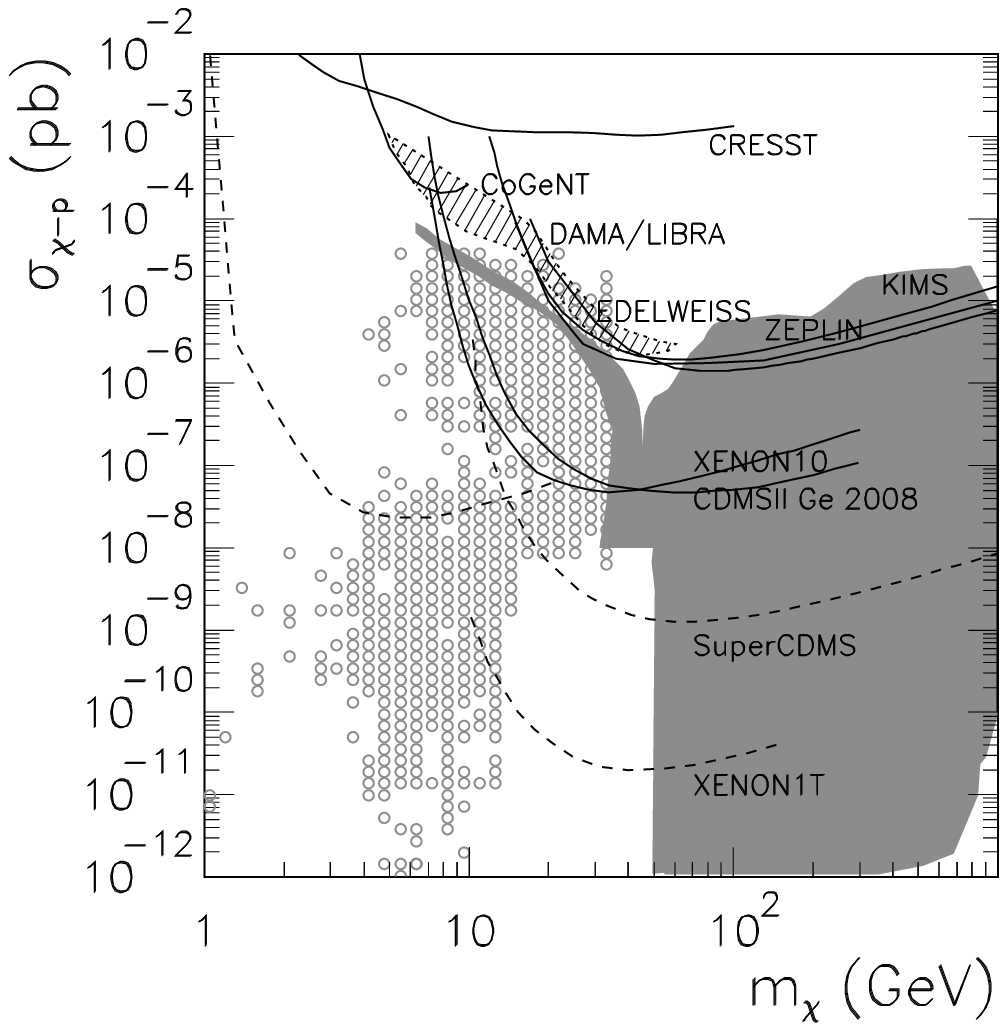}
\hspace*{-0.75cm}
\includegraphics[width=3.1in,angle=0]{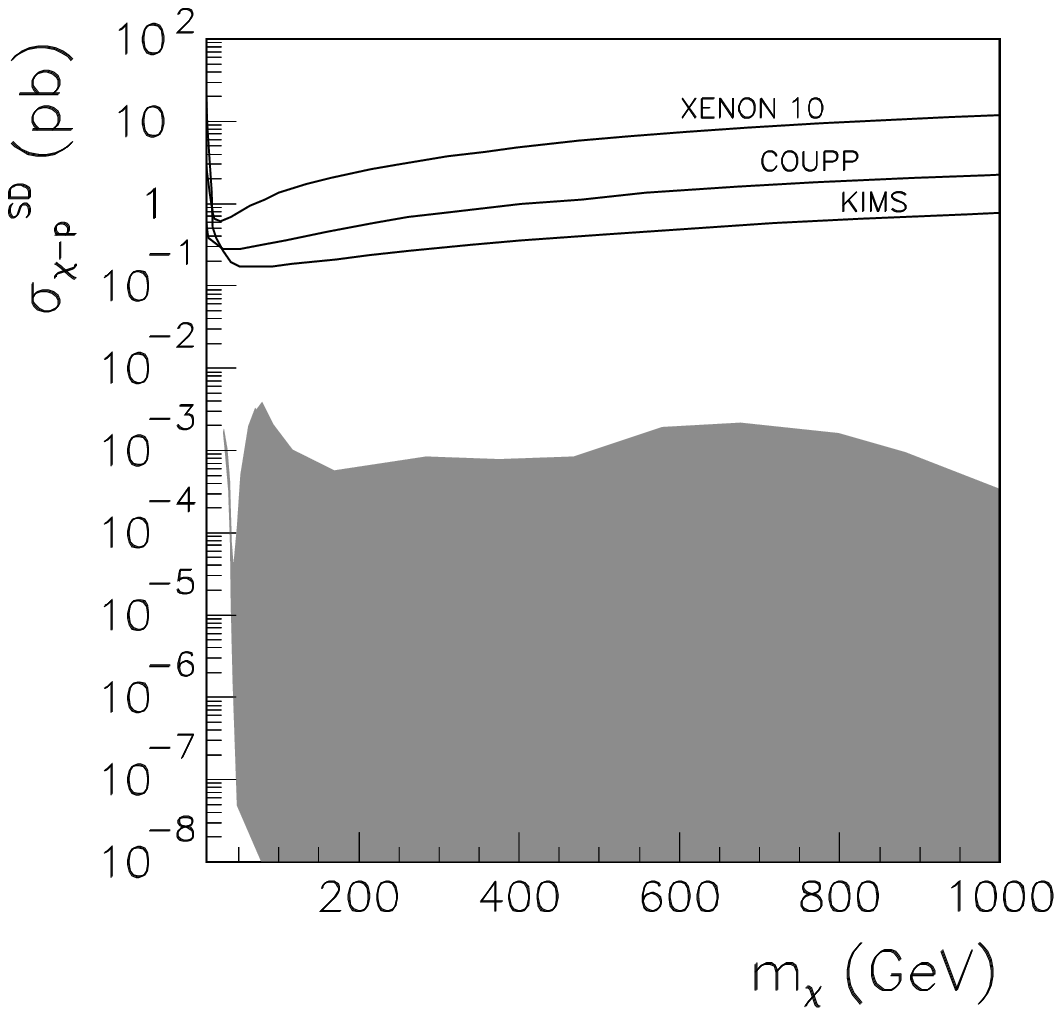}
\caption{Left) 
  Theoretical predictions for neutralino-nucleon SI cross
  section as a function of the neutralino mass obtained by combining the 
  scans of the MSSM parameters from
  Refs.\cite{Kim:2002cy,Bottino:2003cz,Ellis:2005mb,Bottino:2008mf}.  
  The theoretical predictions for the SI cross section of very
  light neutralinos in the NMSSM from Ref.~\cite{Aalseth:2008rx} are
  shown by means of empty grey circles. Present and projected
  experimental sensitivities are displayed using solid and dashed
  lines respectively.
  Right) Theoretical predictions for neutralino-nucleon SD cross
  section as a function of the neutralino mass, using the data
  from the supergravity scan of \cite{Bertone:2007xj}.
} 
\label{fig_neutralino}
\end{figure}

The dependence of the cross sections, and detection prospects, on the
neutralino composition are well known. For example, a large Higgsino
component induces an enhancement of both the Higgs and $Z$ boson
exchange diagrams, thereby leading to an increase in both the SD and
SI cross sections.  On the other hand, the presence of light squarks
(if they become almost degenerate with the neutralino) can lead to an
enhancement of (mainly) the SD cross section.

Analyses of general supersymmetric scenarios with parameters defined
at low energy reveal that the neutralino SI cross section can be as
large as $10^{-5}\pb$ for a wide range of neutralino masses up to
$1\tev$ \cite{Kim:2002cy,Ellis:2005mb}.  Interestingly, when gaugino
masses not fulfilling the GUT relation are allowed, very light
neutralinos with masses $m_{\chione}\gsim7\gev$, and potentially large
cross-sections, can be obtained \cite{Bottino:2002ry,Bottino:2003cz}.
It has been argued that these neutralinos could account for the
DAMA/LIBRA annual modulation signal without contradicting the null
results from CDMS and XENON10 \cite{Bottino:2008mf}, however
this interpretation is now more constrained \cite{Aalseth:2008rx}.
All these features are illustrated in in Fig.\ref{fig_neutralino}.

Analyses have also been done from the point of view of Supergravity
theories, where the SUSY parameters are defined at the GUT scale. In
the simplest case, the CMSSM, the cross-sections are generally small,
since $\chione$ is mostly bino. The largest cross-sections,
$\sigma^{SI} \approx {\cal O}(10^{-8}\pb)$,
are obtained in the focus point region, where the neutralino becomes a
mixed bino-Higgsino state \cite{Edsjo:2004pf,Baer:2005ky}.  
Interestingly, this region seems to be
favoured by recent Bayesian analyses of the CMSSM parameter space
\cite{de-Austri:2006pe,Roszkowski:2007fd,Allanach:2008iq,Trotta:2008bp}.  
Moreover, the predicted SD
cross section is also sizable in the focus point region, approximately
reaching $10^{-4}-10^{-3}\pb$.  In more general Supergravity scenarios the
predicted cross-sections can be significantly larger through the
inclusion of non-universal values for either the scalar masses
\cite{Berezinsky:1995cj,Drees:1996pk,Nath:1997qm,Bottino:1998ka,Arnowitt:1999gq,Accomando:1999eg,Ellis:2002wv,Ellis:2002iu,Drees:2000bs,Ellis:2003eg,Arnowitt:2001ca,Cerdeno:2002rk,Dermisek:2003vn,Baer:2003jb,Baer:2005bu,Ellis:2006jy}
(non-universalities in the Higgs mass parameters being the most
effective), the gaugino masses
\cite{Corsetti:2000yq,Cerdeno:2002rk,Bertin:2002sq,Arnowitt:2002sd,BirkedalHansen:2002am,Chattopadhyay:2003yk},
or both \cite{Pallis:2003aw,Cerdeno:2004zj},

The detection prospects of the lightest neutralino in extended
supersymmetric models may be significantly different, mostly due to
the changes in the Higgs sector and the presence of new neutralino
states.  These constructions are generally referred to as singlet
extensions of the MSSM (see, e.g., \cite{Barger:2007nv}) For example,
this is the case of the NMSSM, in which the presence of very light
Higgses (consistent with LEP constraints if they have a large singlet
composition) can lead to a sizable increase of the SI cross section
\cite{Cerdeno:2004xw,Cerdeno:2007sn,Hugonie:2007vd}. Moreover, 
in the NMSSM very
light neutralinos (with masses below $10 \gev$) are viable
\cite{Gunion:2005rw} and can have very distinctive predictions for
their direct detection, including, for example, smaller SI cross
section than in the MSSM \cite{Aalseth:2008rx}. The theoretical
predictions for the SI cross section of neutralinos with
$m_{\chione}\le 30\gev$ are plotted in Fig.\ref{fig_neutralino}.  In
general, the singlet component of the neutralino does not couple to
the $Z$ boson or to squarks and thus in these constructions the
theoretical predictions for the SD cross section remain the same as in
the MSSM.

Finally, there is another viable supersymmetric WIMP candidate for dark
matter, the lightest sneutrino. The left-handed sneutrino in the MSSM
is excluded given its sizable 
coupling to the $Z$ boson. They therefore either annihilate too rapidly,
resulting in a very small relic abundance, or have large
scattering cross sections which have already been excluded by direct detection
experiments~\cite{Falk:1994es}. 
Several models have been proposed to revive sneutrino DM 
by reducing its coupling with the Z-boson.
This can be achieved by
introducing 
a mixture of left- and right-handed
sneutrinos 
\cite{ArkaniHamed:2000bq,Hooper:2004dc,Arina:2007tm,Arina:2008bb}, 
or by considering a purely right-handed
sneutrino in models with an extended gauge sector \cite{Lee:2007mt} or
Higgs sector
\cite{Garbrecht:2006az,Deppisch:2008bp} such as the NMSSM
\cite{Cerdeno:2008ep}.  
In the first class of models, the elastic scattering of sneutrinos with
quarks would take place
through the $t$-exchange of $Z$ bosons, whereas in the second class it
would mostly be due to the exchange of Higgs bosons. 
The resulting 
SI cross section in these cases can be within the reach
of future detectors for a wide range of sneutrino
masses \cite{Cerdeno:2008ep,Deppisch:2008bp}.
Being a scalar particle, the SD cross section vanishes for
the sneutrino.

\subsection{Kaluza Klein dark matter in UED}

Models of Universal Extra Dimensions, in which all 
fields are allowed to propagate in the bulk \cite{Appelquist:2000nn},
also provide well-motivated candidates for WIMP dark matter in the form 
of the Lightest Kaluza-Klein Particle (LKP), which is likely to be
associated with the first KK excitation  
of the hypercharge gauge boson,
$\bone$~\cite{Cheng:2002iz,Servant:2002aq}.

The elastic scattering of $\bone$ with quarks takes place through the
exchange of KK quarks along $t$ and $s$ channels, which contribute to
both the SD and SI cross section, and a Higgs exchange along a
$t$-channel which only gives a SI contribution
\cite{Servant:2002hb,Cheng:2002ej,Oikonomou:2006mh}.  The theoretical
predictions for the elastic scattering cross sections of $\bone$ are
very dependent on the mass splitting between it and the KK quark,
$\Delta_q$. In particular, both the SI and SD contributions increase
when $\Delta_q$ becomes small, as a consequence of the enhancement of
the contribution from KK quarks.  This is relevant, since in the UED
scenario one expects a quasi-degenerate spectrum, in which the
splittings between the masses are only induced by radiative corrections
\cite{Cheng:2002iz}.  The SI cross section can also be larger in the
presence of light Higgses and for small LKP masses. However the Higgs
mass is generally larger than in the MSSM and its contribution is
suppressed with respect to that of KK quarks.

\begin{figure}
\centering
\includegraphics[width=3.in,angle=0]{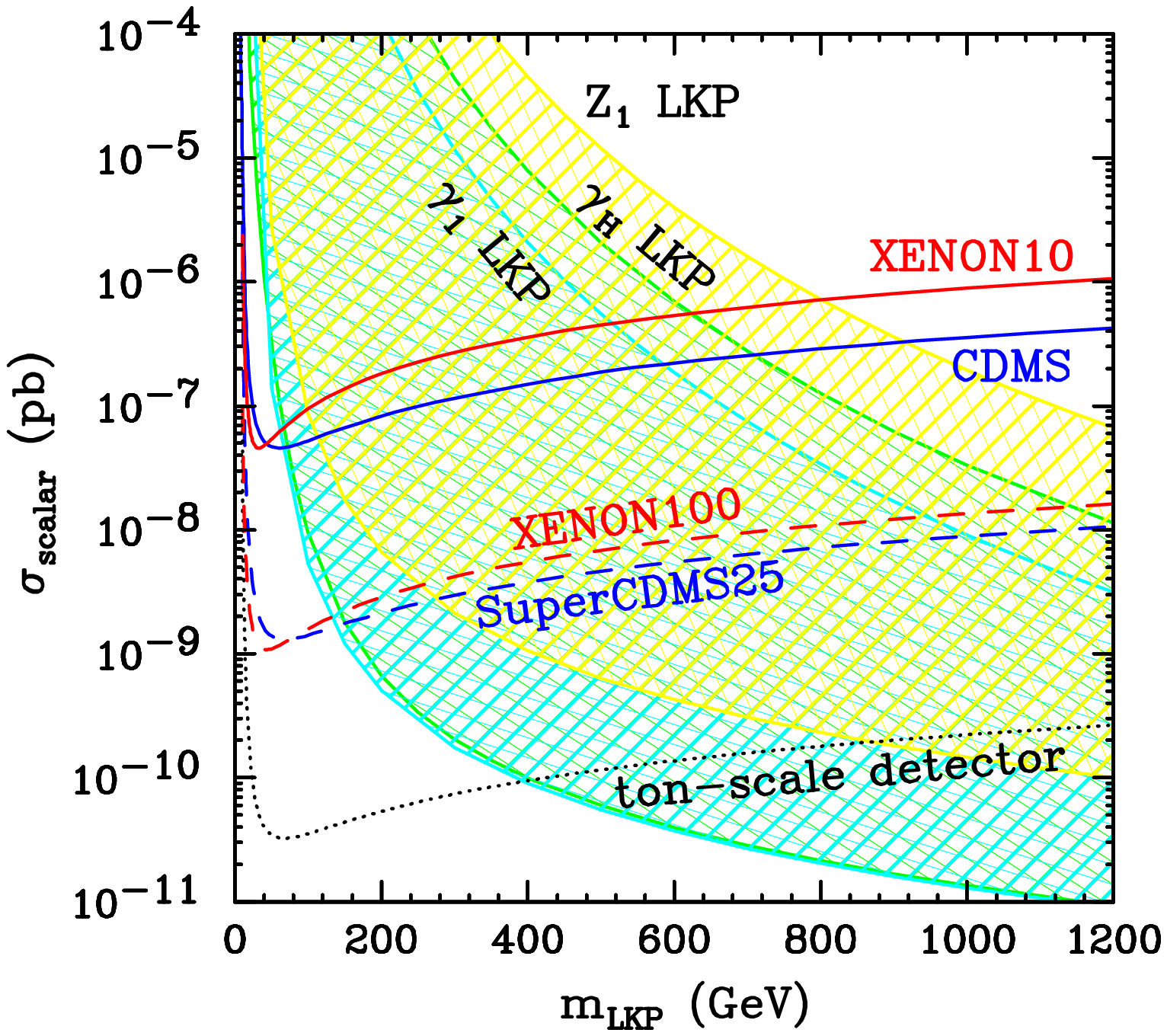}
\includegraphics[width=3.in,angle=0]{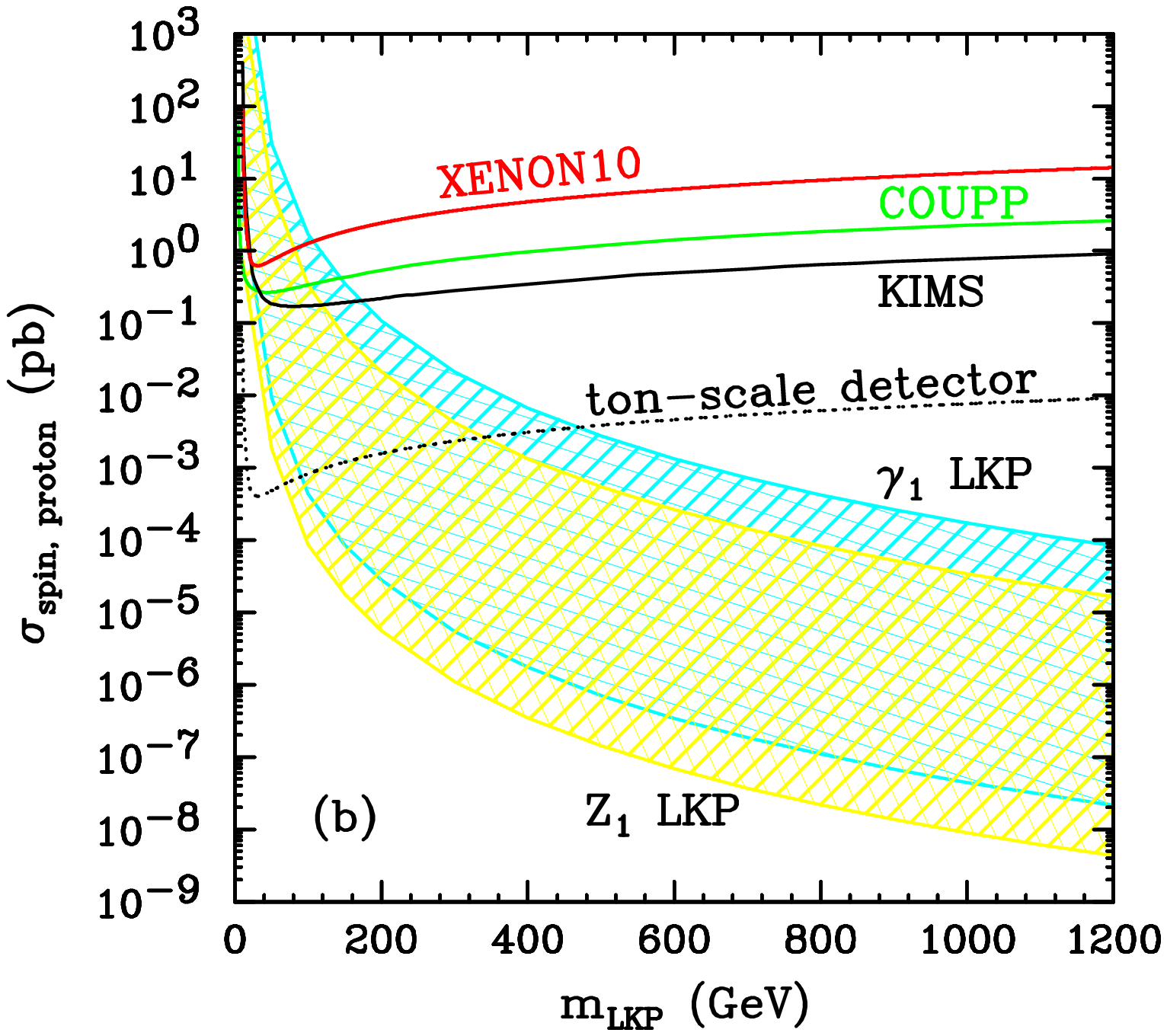}
\caption{Left) Theoretical predictions for the SI LKP-nucleon
  scattering 
  cross section as a function of the LKP mass for the LKP
  candidates discussed in the text. Right) Predictions for the SD
  LKP-proton cross section. Figures extracted from \cite{Arrenberg:2008wy}. }
\label{fig_si_lkp}
\end{figure}

It has been shown that the theoretical predictions for the SI cross
section of $\bone$ can be as large as $\sigma^{SI}\approx10^{-6}$ pb for
masses ranging from $500\,\gev$ to $1\,\tev$, when $\Delta_q\approx0.01$,
for which the correct relic density can be obtained 
\cite{Burnell:2005hm,Kong:2005hn}.
Under the same
conditions, the predicted SD cross section are smaller and, for masses
up to $1$~TeV ton-scale detectors would be required to detect
them. These predictions are illustrated in Fig.\,\ref{fig_si_lkp},
from \cite{Arrenberg:2008wy}.

Other LKP candidates are possible within the UED model if
nonvanishing boundary terms are allowed. More specifically, one may
consider the
first excited states associated with either the $Z$ boson, $\zone$, or
the Higgs, $\hone$ 
\cite{Arrenberg:2008wy}. The detection properties of the $\zone$
are very similar to those of the $\bone$
\cite{Arrenberg:2008wy} (although for the $\zone$ the neutron and
proton spin-dependent cross section are exactly the same, contrary to
the case of the $\bone$).

Finally, in models with two universal extra dimensions the LKP
generally corresponds to the KK excitation associated with the
hypercharge gauge boson, which is called  spinless photon
\cite{Dobrescu:2007ec}. Being a scalar, this particle has no SD cross
section. Its SI cross section can be similar to that of the $\bone$.

\subsection{Little Higgs dark matter}

In these constructions  
a discrete symmetry, called $T$-parity, 
is introduced in order to alleviate the stringent experimental
constraints on low-energy observables. 
A phenomenological consequence of $T$-parity is that the 
Lightest T-odd Particle (LTP) 
becomes absolutely stable. Interestingly, the
LTP is usually the partner of the hypercharge gauge boson
$B_H$
\cite{Cheng:2003ju,Cheng:2004yc,Low:2004xc}.  The LTP-nucleon
scattering cross section receives SI contributions via Higgs and heavy
quark exchange, the latter being the only contribution to the SD part
\cite{Birkedal:2006fz}. Thus, the resulting expressions are very
similar to the case of KK dark matter in UED.  However, unlike the UED
scenario, there is no reason for the heavy quarks to be degenerate in
mass with $A_H$. This, together with the smallness of the heavy quark
Yukawa couplings implies that their scattering cross-sections are in
fact very suppressed.  The SI cross section, being dominated by the
Higgs exchange $t$-channel, increases slightly when the Higgs mass is
small but is nevertheless generally below $10^{-10}\pb$.  The
theoretical predictions for the SD cross section are also very
small. In summary, the direct detection of little Higgs dark matter
is much more difficult than the SUSY and UED cases.

\subsection{Minimal Models and other approaches for dark matter}

Instead of considering DM candidates arising in
existing theories beyond the SM, a bottom-up approach can be adopted
in which minimal additions to the SM are considered, involving
the inclusion of a WIMP field (usually a new singlet) and
new symmetries that protect their decay (in some cases, also a new
``mediator'' sector that couples the WIMP to the SM). 
Examples in this direction include WIMPs with singlet mediation
\cite{McDonald:1993ex,Burgess:2000yq,Lisanti:2007ec}, 
models with an extended electroweak sector
\cite{Cirelli:2005uq,Barbieri:2006dq,LopezHonorez:2006gr}, 
models with additional gauge groups, and the 
Secluded Dark Matter scenario \cite{Pospelov:2007mp} in which WIMPs
could escape direct detection.

The theoretical predictions for the direct detection of WIMPs in this
class of models are very dependent on the mediator sector, since it
determines the couplings of WIMPs to ordinary matter (quarks). 
For example, scalar WIMPs interact with ordinary quarks through the
exchange of Higgs bosons in a $t$ channel
\cite{McDonald:1993ex,Burgess:2000yq}. The singlet coupling to the
Higgs is constrained in order to reproduce the correct relic
abundance, thus leaving only the WIMP and Higgs masses as free
parameters. The resulting cross-sections increase as both masses
decrease and can be as large as $10^{-6}\pb$ for very light WIMPs of
order $10\gev$ but are reduced to be below $10^{-8}\pb$ when the DM
candidate has a mass above $100\gev$.

Other scenarios can have more general couplings.  In the Inert Doublet
Model, where elastic scattering proceeds only through the exchange of a
Higgs or a $Z$ boson along a $t$ channel, the resulting SI cross
section is rather small. Only for light WIMPs, with masses below
$100\gev$, is the predicted SI cross section large enough to be
experimentally tested (from about $10^{-10}\pb$ to as much as
$10^{-7}\pb$) whereas the predictions for heavy WIMPs are well below
the sensitivity of ton-size experiments, usually of order
$10^{-13}\pb$ \cite{LopezHonorez:2006gr}.

In the Minimal DM approach of \cite{Cirelli:2005uq} WIMP candidates
with direct couplings to the $Z$ boson are already excluded by direct
DM searches. However, some fermionic candidates are still viable which
interact with quarks through the exchange of $W$ (and Higgs)
bosons. These particles (with masses of several TeV) can have an SI
cross section of order $10^{-8}\pb$.

\subsection{Inelastic cross section}

It is finally worth mentioning that the WIMP-nucleon cross section can
also receive a contribution from inelastic scattering by creating either
an excited nuclear \cite{Ellis:1988nb}
or electronic state \cite{Starkman:1995ye} or even by creating an
excited WIMP state  
\cite{TuckerSmith:2001hy,TuckerSmith:2004jv,Chang:2008gd}. 
The last possibility is particularly interesting if the mass
difference, $\delta$, between the excited dark
matter candidate, $\chi_2$ and WIMP $\chi_1$ is of order of the WIMP
kinetic energy (i.e., about $100\kev$). In that case, the inelastic
scattering off nuclei $\chi_1 N \to \chi_2 N$ can occur and the only
kinematic change is in the minimal WIMP velocity that can trigger a
specific recoil energy, which is 
increased by 
$\Delta v_{min}=\delta/\sqrt{(2\mnucl E_R)}$.
This clearly favours detection in heavy targets such as Iodine (since
$\Delta v_{min}$ is smaller) 
and might provide a possible
explanation for the DAMA/LIBRA signal compatible with the
null restuls in other experiments \cite{Chang:2008gd}.

\subsection{Discrimination of Dark Matter candidates}

As illustrated by figs.~\ref{fig_neutralino} and \ref{fig_si_lkp},
current experiments are already probing the masses and cross-sections
predicted for various WIMP candidates. Furthermore future experiments
will be sensitive to a substantial fraction of the parameter space. If
any of these experiments succeed in detecting dark matter particles,
the next objective will be to identify its particle nature.

\begin{figure}
\centering
\includegraphics[width=6in,angle=0]{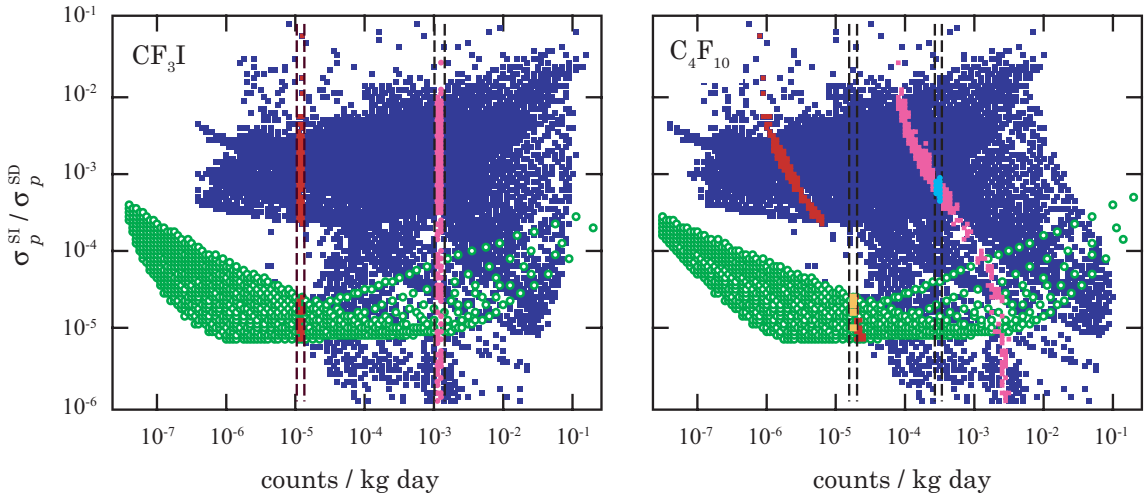}
\caption{Left) The detection with a given target such as  
  CF$_3$I 
  can only loosely constrain models for the dark matter (blue squares
  for neutralinos, green circles for the LKP) in the
  $\sigma^{\rm SI}_p / \sigma^{\rm SD}_p$ versus count-rate
  plane. Red (magenta) dots show the many models consistent with a
  measurement of  $\sim10^{-5}$ ($10^{-3}$) counts/kg day. 
  Right) measurement of the event rate in a
  second target such as C$_4$F$_{10}$, with lower
  sensitivity to spin-independent couplings, effectively reduces the
  remaining number of allowed models--orange (aqua) dots--and
  generally allows discrimination between the neutralino and the
  LKP. Figures extracted from \cite{Bertone:2007xj}. }
\label{fig_discrimination}
\end{figure}

In this sense, the simultaneous measurement of both the SI and SD dark
matter couplings, through experiments which are sensitive to both
signals, can provide very valuable information \cite{Bertone:2007xj}.
This is illustrated in Fig.~\ref{fig_discrimination}, where the ratio
of SI to SD cross section is plotted for the neutralino and LKP cases
versus the event rate for two complementary choices of target
materials (that could be used in the COUPP experiment).  As shown in
the left panel, the measurement of an event rate in a single detector
does reduce the number of allowed models, but does not generally place
significant constraints on coupling parameters or on the nature of
the dark matter detected. However, as shown in the right panel, a
subsequent detection using a second complementary target
does substantially reduce the allowed range of coupling parameters,
and allows, in most cases, an effective discrimination between
neutralino and LKP candidates.

This analysis can be extended to other dark matter candidates.
Moreover, in order to eliminate the large astrophysical and
theoretical uncertainties which affect the dark matter rates, the
ratio of WIMP-proton and WIMP-neutron amplitudes can be used
\cite{Belanger:2008gy} and compared for different target materials.
For example, the comparative study of the ratios of SD neutron to
proton amplitudes can provide a good discrimination of dark matter
models by distinguishing candidates for which the SD cross section is
dominated by $Z$ boson exchange (such as the neutralino in some
regions of the parameter space) from those where the dominant channel
is squark or KK quark exchange (such as the LKP or LTP).  A similar
analysis for the SI neutron to proton ratio can be used to disentangle
models with dominant Higgs or $Z$ boson exchange, however in practise
the different target materials are less sensitive to these
differences.  Finally, the mass determination techniques described in
the Sec.~\ref{sec:drde} can provide complementary information that could
lead to more effective discrimination between the various dark matter
models.

\paragraph{Acknowledgements.}
The authors would like to thank Daniele Fantin, Michael Merrifield,
and Ben Morgan for useful discussions and comments. DGC is supported
by  the ``Ram\'on y Cajal'' program of the
Spanish MICINN, by the Spanish grants FPA2009-08958, 
HEPHACOS S2009/ESP-1473, and  MICINN's Consolider-Ingenio
2010 Program MULTIDARK CSD2009-00064, and by the EU network
PITN-GA-2009-237920.
AMG is supported by STFC.


\end{document}